\documentclass[preprint,12pt]{elsarticle}
\usepackage{amsmath}
\usepackage{amssymb}
\usepackage[pdftex,colorlinks]{hyperref}
\usepackage{multirow}
\usepackage{enumerate}
\usepackage{cases}

\makeatletter

\newcommand {\Rmnum} [1] {\expandafter \@slowromancap \romannumeral #1@}
\makeatother

\newtheorem{Protocol}{Protocol}

\biboptions{compress}

\begin{document}

\begin{frontmatter}

\title{Quantum no-key protocol for secure communication of classical message}


\author{Li Yang}
\address{State Key Laboratory of Information Security,
 Institute of Information Engineering, Chinese Academy of Sciences, Beijing 100093, China}

\begin{abstract}
We propose a class of quantum no-key protocols for private communication of classical message based on quantum computing of random Boolean permutations, and demonstrate that they are information-theoretic secure. These protocols are designed to resist middleman attack for two parties preshared authentication key, and achieve perfect mutual data origin authentication which ensures the permanent reusing of authentication key. Finally, we simplify the protocol to a 4-round one, and show that any protocol with 3 or less rounds cannot achieve perfect security without consuming preshared key.
\end{abstract}


\end{frontmatter}


Quantum no-key protocol is one of the earliest interactive quantum secure communication protocols\cite{YW02,YWL02,BF02}, which remains three-round structure of Shamir's original idea, but can be improved to resist man-in-the-middle (MIM) attack\cite{YWL02,Y03,YL12}. This kind of protocols have been applied to meet various cryptographic demands\cite{JL12,KSM05,K06,K05}. When the communication is limited to transmitting classical message, and we do not require it to keep exponential-security under the MIM attack, a protocol can be simplified to a two-round one\cite{FLS03}. So far, there are various quantum no-key protocols presented \cite{YL12}, but none with rigorous security proof. Here we propose a new quantum no-key protocol for classical message communication with provable perfect security.

\section{Basic protocol}
Alice intends to transmit classical message $x$ to Bob through a quantum channel. She can complement it by the following protocol:
\begin{Protocol}

~~~~~~~~~~~\emph{\begin{enumerate}[(1)]
\item Alice prepares quantum state:
\begin{eqnarray}
 |x\rangle_{\Rmnum{1}}&\overset{{H^{(n)}}}\longrightarrow&\frac{1}{\sqrt{2^{n}}}\sum\limits_{m}(-1)^{x\cdot m}|m\rangle_{\Rmnum{1}}.
 \end{eqnarray}
\item Alice chooses a Boolean permutation $F_{A}(\cdot)$ randomly, and executes:
\begin{eqnarray}
\frac{1}{\sqrt{2^{n}}}\sum\limits_{m}(-1)^{x\cdot m}|m\rangle_{\Rmnum{1}}|0\rangle_{\Rmnum{2}}&\rightarrow&\frac{1}{\sqrt{2^{n}}}\sum\limits_{m}(-1)^{x\cdot m}|m\rangle_{\Rmnum{1}}|F_{A}(m)\rangle_{\Rmnum{2}}.
 \end{eqnarray}
\item Alice sends quantum register $\Rmnum{1}$ to Bob, the state appeared in the channel is:
\begin{eqnarray}\nonumber
\rho_{1}&=&tr_{\Rmnum{2}}\left[\frac{1}{2^{n}}\sum\limits_{m, n}(-1)^{x\cdot m}(-1)^{x\cdot n}|m\rangle_{\Rmnum{1}}\langle n|\otimes |F_{A}(m)\rangle_{\Rmnum{2}}\langle F_{A}(n)|\right]\\
&=&\frac{1}{2^{n}}\sum\limits_{m, n}(-1)^{x\cdot m}(-1)^{x\cdot n}|m\rangle_{\Rmnum{1}}\langle n|\sum\limits_{k}\langle k| F_{A}(m)\rangle\langle F_{A}(n)|k\rangle.
 \end{eqnarray}
 Since $F_{A}(\cdot)$ is a Boolean permutation, we have:
\begin{eqnarray}
\rho_{1}&=&\frac{1}{2^{n}}\sum\limits_{m}|m\rangle\langle m|=\frac{1}{2^{n}}I_{2^{n}}.
\end{eqnarray}
That is, the state appeared on the channel is only an ultimate mixed state.
\item After receiving the quantum state $\rho_{1}$, Bob randomly chooses a Boolean permutation $F_{B}(\cdot)$ and computes:
   \begin{eqnarray}
|m\rangle_{\Rmnum{1}}|0\rangle_{\Rmnum{3}}&\rightarrow&|m\rangle_{\Rmnum{1}}|F_{B}(m)\rangle_{\Rmnum{3}}.
\end{eqnarray}
Then, the state of quantum register $\Rmnum{1}$ will be:
  \begin{eqnarray}\nonumber
\rho_{2}&=&tr_{\Rmnum{3}}\left[\frac{1}{2^{n}}\sum\limits_{m}|m\rangle_{\Rmnum{1}}\langle m|\otimes|F_{B}(m)\rangle_{\Rmnum{3}}\langle F_{B}(m)|\right]\\\nonumber
&=&\frac{1}{2^{n}}\sum\limits_{m}|m\rangle_{\Rmnum{1}}\langle m|\otimes\sum\limits_{k}\langle k|F_{B}(m)\rangle_{\Rmnum{3}}\langle F_{B}(m)|k\rangle_{\Rmnum{3}}\\
&=&\frac{1}{2^{n}}I_{2^{n}}.
\end{eqnarray}
That is, while Bob sending the quantum register $\Rmnum{1}$ to Alice, the state in the channel is also an ultimate mixed state.
\item After received register $\Rmnum{1}$, Alice does computation:
 \begin{eqnarray}\nonumber
|m\rangle_{\Rmnum{1}}|F_{A}(m)\rangle_{\Rmnum{2}}&\rightarrow&|m\rangle_{\Rmnum{1}}|F_{A}(m)\oplus F_{A}(m)\rangle_{\Rmnum{2}}\\
&=&|m\rangle_{\Rmnum{1}}|0\rangle_{\Rmnum{2}},
\end{eqnarray}
and sends register $\Rmnum{1}$ to Bob again. Then the state transmitted in the channel will be:
\begin{eqnarray}
\rho_{3}&=&tr_{\Rmnum{2}}\left[\frac{1}{2^{n}}\sum\limits_{m}|m\rangle_{\Rmnum{1}}\langle m||0\rangle_{\Rmnum{2}}\langle 0|\right]=\frac{1}{2^{n}}I_{2^{n}}.
\end{eqnarray}
The transmitted state is an ultimate mixed state again.
\item After received $\rho_{3}$, Bob executes computation:
\begin{eqnarray}\nonumber
|m\rangle_{\Rmnum{1}}|F_{B}(m)\rangle_{\Rmnum{2}}&\rightarrow&|m\rangle_{\Rmnum{1}}|F_{B}(m)\oplus F_{B}(m) \rangle_{\Rmnum{2}}\\
&=&|m\rangle_{\Rmnum{1}}|0\rangle_{\Rmnum{2}},
\end{eqnarray}
and obtains the state: $\frac{1}{\sqrt{2^{n}}}\sum\limits_{m}(-1)^{x\cdot m}|m\rangle_{\Rmnum{1}}$. Finally, Bob has the message $x$ via performing $H^{(n)}$ transformation and measuring register $\Rmnum{1}$.
\end{enumerate}}
\end{Protocol}
It can be seen more clearly if we exhibit the whole evolution of $(A, B)$ composite system:
 \begin{eqnarray}\nonumber
&&\frac{1}{\sqrt{2^{n}}}\sum\limits_{m}(-1)^{x\cdot m}|m\rangle_{\Rmnum{1}}|0\rangle_{\Rmnum{2}}|0\rangle_{\Rmnum{3}}\\\nonumber
&\rightarrow& \frac{1}{\sqrt{2^{n}}}\sum\limits_{m}(-1)^{x\cdot m}|m\rangle_{\Rmnum{1}}|F_{A}(m)\rangle_{\Rmnum{2}}|0\rangle_{\Rmnum{3}}\\\nonumber
&\rightarrow&\frac{1}{\sqrt{2^{n}}}\sum\limits_{m}(-1)^{x\cdot m}|m\rangle_{\Rmnum{1}}|F_{A}(m)\rangle_{\Rmnum{2}}|F_{B}(m)\rangle_{\Rmnum{3}}\\\nonumber
\end{eqnarray}
\begin{eqnarray}\nonumber
&\rightarrow&\frac{1}{\sqrt{2^{n}}}\sum\limits_{m}(-1)^{x\cdot m}|m\rangle_{\Rmnum{1}}|F_{A}(m)\oplus |F_{A}(m) \rangle_{\Rmnum{2}}|F_{B}(m)\rangle_{\Rmnum{3}}\\\nonumber
&=&\frac{1}{\sqrt{2^{n}}}\sum\limits_{m}(-1)^{x\cdot m}|m\rangle_{\Rmnum{1}}|0 \rangle_{\Rmnum{2}}|F_{B}(m)\rangle_{\Rmnum{3}}\\\nonumber
&\rightarrow&\frac{1}{\sqrt{2^{n}}}\sum\limits_{m}(-1)^{x\cdot m}|m\rangle_{\Rmnum{1}}|0 \rangle_{\Rmnum{2}}|F_{B}(m)\oplus F_{B}(m) \rangle_{\Rmnum{3}}\\
&=&\frac{1}{\sqrt{2^{n}}}\sum\limits_{m}(-1)^{x\cdot m}|m\rangle_{\Rmnum{1}}|0 \rangle_{\Rmnum{2}}|0 \rangle_{\Rmnum{3}}.
\end{eqnarray}
That is, Bob obtains the quantum state of the first quantum register: $$\frac{1}{\sqrt{2^{n}}}\sum\limits_{m}(-1)^{x\cdot m}|m\rangle_{\Rmnum{1}}.$$.

With respect to secrecy, $F_{A}(\cdot), F_{B}(\cdot)$ are both randomly selected Boolean permutation, and the transmitted quantum states $\rho_{1}, \rho_{2},\rho_{3}$ are all ultimate mixed states, so the adversary can not get any information, protocol 1 has information-theoretic security.
However, this protocol cannot resist MIM attack. When that the adversary Eve impersonates Bob to communicate with Alice, at the same time, he impersonates Alice to communicate with Bob, he will get $x$ from Alice and send $x^{'}$ to Bob successfully.

\section{Protocol with data origin authentication}
In this section, we improve protocol 1 to be one with mutual authentication. The protocol we shall construct is a 9-round no-key protocol with mutual identification. The whole protocol consists of three stages, each includes three rounds.
\begin{Protocol}

~~~~~~~~~~~~~~~~~~~~~~~~~~~~~~~~~~~

\emph{Alice and Bob share randomly chosen Boolean functions $s_{A}(\cdot)$ and $s_{B}(\cdot)$ in advance.}
\emph{\begin{enumerate}[(1)]
\item Alice randomly chooses Boolean permutation $F_{A}(\cdot)$ and random number $r_{A}$, then prepares quantum state:
\begin{eqnarray}
\sum\limits_{m}\alpha_{m}|m\rangle_{\Rmnum{1}}|0\rangle_{\Rmnum{2}}|0\rangle_{\Rmnum{3}}
&\rightarrow&\sum\limits_{m}\alpha_{m}|m\rangle_{\Rmnum{1}}|F_{A}(m)\rangle_{\Rmnum{2}}|s_{A}(m)\oplus r_{A}\rangle_{\Rmnum{3}}
\end{eqnarray}
where $\alpha_{m}=\frac{1}{\sqrt{2^{n}}}(-1)^{x\cdot m}$, then Alice sends the first and the third registers to Bob.
For the adversary, the quantum state transmitted  in the channel is:
\begin{footnotesize}
\begin{eqnarray}\nonumber
\sigma_{1}&=&tr_{\Rmnum{2}}\left[\sum\limits_{m, n, r_{A}}\frac{\alpha_{m}\alpha^{*}_{n}}{2^{l}}|m\rangle\langle n|\otimes|F_{A}(m)\rangle\langle F_{A}(n)|\otimes|s_{A}(m)\oplus r_{A}\rangle\langle s_{A}(n)\oplus r_{A}|\right]\\\nonumber
&=&\frac{1}{2^{l}}\sum\limits_{m, r_{A}}|\alpha_{m}|^{2}|m\rangle\langle m|\otimes|s_{A}(m)\oplus r_{A}\rangle\langle s_{A}(m)\oplus r_{A}|\\\nonumber
&=&\frac{1}{2^{l}}\sum\limits_{m}|\alpha_{m}|^{2}|m\rangle\langle m|\otimes I_{2^{l}}\\\nonumber
&=&\frac{1}{2^{l+n}}I_{2^{n}}\otimes I_{2^{l}}\\
&=&\frac{1}{2^{n+l}}I_{2^{n+l}}
\end{eqnarray}
\end{footnotesize}
Therefore, $\sigma_{1}$ is an ultimate mixed state.
\item Bob randomly chooses Boolean permutation $F_{B}(\cdot)$ and random number $r_{B}$, then uses $F_{B}(\cdot)$, $r_{B}$ and the preshared Boolean function $s_{B}(\cdot)$ to accomplish computation:
  \emph{\begin{eqnarray}\nonumber
&&|m\rangle_{\Rmnum{1}}|s_{A}(m)\oplus r_{A}\rangle_{\Rmnum{3}}|0\rangle_{\Rmnum{4}}|0\rangle_{\Rmnum{5}}\\
&\rightarrow&|m\rangle_{\Rmnum{1}}|r_{A}\rangle_{\Rmnum{3}}|F_{B}(m)\rangle_{\Rmnum{4}}|s_{B}(m)\oplus r_{B}\rangle_{\Rmnum{5}}.
\end{eqnarray}}
~~Then Bob measures the third quantum register and sends the first and the fifth register to Alice. Similar to the analysis in step $(1)$, we can see that quantum state $\sigma_{2}$ in the channel is also an ultimate mixed state.
\item Alice chooses random number $r_{A^{'}}$ and do the transformation:
  \begin{eqnarray}\nonumber
&&|m\rangle_{\Rmnum{1}}|F_{A}(m)\rangle_{\Rmnum{2}}|s_{B}(m)\oplus r_{B}\rangle_{\Rmnum{5}}|0\rangle_{\Rmnum{6}}\\\nonumber
&\rightarrow&|m\rangle_{\Rmnum{1}}|F_{A}(m)\oplus F_{A}(m)\rangle_{\Rmnum{2}}|s_{B}(m)\oplus s_{B}(m)\oplus r_{B}\rangle_{\Rmnum{5}}|s_{A}(m)\oplus r_{A^{'}}\rangle_{\Rmnum{6}}\\
&=&|m\rangle_{\Rmnum{1}}|0\rangle_{\Rmnum{2}}|r_{B}\rangle_{\Rmnum{5}}|s_{A}(m)\oplus r_{A^{'}}\rangle_{\Rmnum{6}}.
\end{eqnarray}
Then Alice measures the fifth quantum register, and sends register $\Rmnum{1}$ and register $\Rmnum{6}$ to Bob. It can also be proved that the quantum state appeared in channel is $\sigma_{3}=\frac{1}{2^{n+l}}I_{2^{n+l}}$.
\item After he receives $\sigma_{3}$, Bob executes transformation:
 \begin{eqnarray}\nonumber
&&|m\rangle_{\Rmnum{1}}|F_{B}(m)\rangle_{\Rmnum{4}}|s_{A}(m)\oplus r^{'}_{A}\rangle_{\Rmnum{6}}\\\nonumber
&\rightarrow&|m\rangle_{\Rmnum{1}}|F_{B}(m)\oplus F_{B}(m)\rangle_{\Rmnum{4}}|s_{A}(m)\oplus s_{A}(m)\oplus r^{'}_{A}\rangle_{\Rmnum{6}}\\
&=&|m\rangle_{\Rmnum{1}}|0\rangle_{\Rmnum{4}}|r^{'}_{A}\rangle_{\Rmnum{6}},
\end{eqnarray}
and obtains the quantum state : $\frac{1}{\sqrt{2^{n}}}\sum\limits_{m}(-1)^{x\cdot m}|m\rangle$.
\item Bob performs Hadamard transformation on the received quantum state, and measures the quantum register $\Rmnum{1}$. Finally, he obtains the classical message $x$.
\end{enumerate}}
\end{Protocol}
~~~~There exists an obvious weakness in the above protocol. Though Eve cannot obtain $x$, she can tamper the system without being detected. A direct way to resist this attack is to execute the protocol three times as follows:
 \begin{Protocol}
\emph{\begin{enumerate}[(1)]
\item The first stage (1-3 rounds): Alice sends classical message $x$ to Bob via executing protocol 2. We can see that, for Eve without $s_{A}(\cdot)$ and $s_{B}(\cdot)$, obtaining the message $x$ is a too difficult task to accomplish.
\item The second stage (4-6 rounds): Bob sends the message $x$ back to Alice via protocol 2. We can see that even if Eve impersonates Bob to communicate with Alice, she cannot send back the right $x$ to Alice, and then Alice will find that she has not accomplished the communication task.
\item The third stage (7-9 rounds): Alice sends $x$ to Bob again via protocol 2. If Eve impersonates Alice to communicate with Bob, Bob will receive two different messages in this and the first stage respectively, because Eve does not have $s_{A}$ and $s_{B}$. Thererfore, via this stage, Bob can discriminate the attacker from the expected message sender Alice.
\end{enumerate}}
\end{Protocol}
~~~~The interaction of this protocol is up to 9 so as to achieve perfect security. Since all the quantum states transmitted are ultimate mixed states, the adversary cannot get any information.
\begin{enumerate}[(1)]
\item For different message $x$ and $y$, the quantum states in the channel are all the ultimate mixed state, so the trace distance of $\rho_{i}(x)$ and $\rho_{j}(y)$ $(i= 1,2,3)$ is:
  \begin{eqnarray}
D(\rho_{i}(x), \rho_{j}(y))=0
\end{eqnarray}
Therefore, Eve cannot attack the plaintext directly.
\item For Alice and Bob's authentication Boolean function $s_{A}$, $s_{B}$, due to $r_{A}$ and $r_{B}$ are local random bit strings chosen for each round independently, we have:
    \begin{footnotesize}
    \begin{eqnarray}
D(\rho_{i}(x, s_{A}(\cdot)), \rho_{j}(y, s^{'}_{A}(\cdot))=D(\rho_{i}(x, s_{B}(\cdot)), \rho_{j}(y, s^{'}_{B}(\cdot))=D(\frac{I_{2^{n+l}}}{2^{n+l}}, \frac{I_{2^{n+l}}}{2^{n+l}})=0
\end{eqnarray}
\end{footnotesize}
That is, the adversary cannot attack $s_{A}$, $s_{B}$.


~~~~Though this 9-round protocol is constructed so complicated, there still exists a MIM attack. The adversary can perform $Z^{x}$ operation to quantum register $\Rmnum{1}$ in each of the 9 passes: $Z^{x^{'}}|m\rangle=(-1)^{x^{'}\cdot m}|m\rangle$. This attack will change the final message Bob recieved from $x$ to $x\oplus x^{'}$. We can easily understand this attack from the basic relation $HX=ZH$.

~~~~To resist this attack, Alice and Bob have to use authentication key to code $x$: $x\rightarrow e_{k}(x)$. While Eve tampered, Alice and Bob will find that the message received has been changed. That is, when we use this additional authentication, Eve cannot implement her MIM attack without being detected.
\end{enumerate}
\section{Simplification of the previous protocols}

The protocol 2 can be simplified as follows:
\begin{Protocol}
\emph{\begin{enumerate}[(1)]
\item The message receiver Bob executes computation as below:
\begin{eqnarray}\nonumber
|0\rangle_{\Rmnum{1}}|0\rangle_{\Rmnum{2}}|0\rangle_{\Rmnum{3}}&\rightarrow&
\frac{1}{\sqrt{2^{n}}}\sum\limits_{m}|m\rangle_{\Rmnum{1}}|0\rangle_{\Rmnum{2}}|0\rangle_{\Rmnum{3}}\\\nonumber
&\rightarrow&\frac{1}{\sqrt{2^{n}}}\sum\limits_{m}|m\rangle_{\Rmnum{1}}|F_{B}(m)\rangle_{\Rmnum{2}}|0\rangle_{\Rmnum{3}}\\
&\rightarrow&\frac{1}{\sqrt{2^{n}}}\sum\limits_{m}|m\rangle_{\Rmnum{1}}|F_{B}(m)\rangle_{\Rmnum{2}}|s_{B}(m)\oplus r_{B}\rangle_{\Rmnum{3}},
\end{eqnarray}
where $F_{B}(\cdot)$ is a randomly chosen Boolean permutation, $s_{B}$ is a Boolean Function preshared by Alice and Bob for authentication, and $r_{B}$ is a randomly chosen bit string. Then, Bob sends register $\Rmnum{1}$ and $\Rmnum{3}$ together to Alice.
\item The message sender Alice executes computation:
\begin{eqnarray}\nonumber
|m\rangle_{\Rmnum{1}}|s_{B}(m)\oplus r_{B}\rangle_{\Rmnum{3}}&\rightarrow&|m\rangle_{\Rmnum{1}}|s_{B}(m)\oplus s_{B}(m)\oplus r_{B}\rangle_{\Rmnum{3}}\\
&=&|m\rangle_{\Rmnum{1}}|r_{B}\rangle_{\Rmnum{3}}
\end{eqnarray}
and measures the third register , then she performs quantum operation $Z^{x}$ to register $\Rmnum{1}$: $Z^{x}|m\rangle_{\Rmnum{1}}=(-1)^{x\cdot m}|m\rangle_{\Rmnum{1}}$, and executes computation: $|m\rangle_{\Rmnum{1}}|0\rangle_{\Rmnum{4}}\rightarrow|m\rangle_{\Rmnum{1}}|s_{A}(m)\oplus r_{A}\rangle_{\Rmnum{4}}$.
Then she sends the two registers to Bob. Where $s_{A}$ is another Boolean function preshared by Alice and Bob for authentication, and $r_{A}$ is a bit-string randomly chosen by Alice.
\item Bob computes:
\begin{eqnarray}\nonumber
|m\rangle_{\Rmnum{1}}|s_{A}(m)\oplus r_{A}\rangle_{\Rmnum{4}}&\rightarrow&|m\rangle_{\Rmnum{1}}|s_{A}(m)\oplus s_{A}(m)\oplus r_{A}\rangle_{\Rmnum{4}}\\
&=&|m\rangle_{\Rmnum{1}}|r_{A}\rangle_{\Rmnum{4}},
\end{eqnarray}
and measures the register $\Rmnum{4}$, then he computes:
\begin{eqnarray}\nonumber
|m\rangle_{\Rmnum{1}}|F_{B}(m)\rangle_{\Rmnum{2}}&\rightarrow&|m\rangle_{\Rmnum{1}}|F_{B}(m)\oplus F_{B}(m)\rangle_{\Rmnum{2}}\\
&=&|m\rangle_{\Rmnum{1}}|0\rangle_{\Rmnum{2}},
\end{eqnarray}
and obtains the quantum state:
\begin{eqnarray}
\frac{1}{\sqrt{2^{n}}}\sum\limits_{m}(-1)^{x\cdot m}|m\rangle_{\Rmnum{1}}&\overset{{H^{(n)}}}\longrightarrow&|x\rangle_{\Rmnum{1}}
\end{eqnarray}
Finally, Bob measures the register $\Rmnum{1}$ and results in the message $x$ from Alice.
\end{enumerate}}
\end{Protocol}
Because $F_{B}(\cdot)$ is a Boolean permutation, we can prove as in protocol 2 that the two quantum states transmitting in channel are both ultimate mixed states.
Based on this results, the protocol 3 can be simplified to a 6-round protocol with mutual authentication:
\begin{Protocol}
\emph{\begin{enumerate}[(1)]
\item Alice executes protocol 4 to send a message $x$ to Bob;
\item Bob executes protocol 4 to send $x$ back to Alice;
\item Alice executes protocol 4 with Bob to send $x$ to Bob again.
\end{enumerate}}
\end{Protocol}
Via this three stage protocol, Alice can confirm that the message $x$ is really received by the expected receiver Bob, and Bob also acknowledges that the message is really from the expected sender Alice. Actually, this protocol can be simplified further to a 4-round one as follows:
\begin{Protocol}
\emph{\begin{enumerate}[(1)]
\item Alice encodes: $x\rightarrow e_{k}(x)$ with some information-theoretic secure authentication code, and executes the first step of protocol 5 to send $x^{'}=(x, e_{k}(x))$ to Bob.
\item Bob executes the second step of protocol 5, and sends $e_{k}(x)$ back to Alice.
\end{enumerate}}
\end{Protocol}
~~~~It can be seen that since a classical authentication code $e_{k}$ is embedded in the protocol, the 6-round protocol can be simplified to a 4-round one which can resist the MIM attack with operation $Z^{x}$.


\section{Security analysis of protocols with 3 or less rounds}
\textbf{A. Non-interactive protocol}

The sender Alice intends to send $x$ to receiver Bob, she firstly computes $e_{k}(x)$, the MAC of \emph{x}, and gets $x^{'}=(x, e_{k}(x))$. Then she computes as below:
\begin{eqnarray}\nonumber
|x^{'}\rangle&\rightarrow&\frac{1}{\sqrt{2^{n}}}\sum\limits_{m}(-1)^{x^{'}\cdot m}|m\rangle_{\Rmnum{1}}|0\rangle_{\Rmnum{2}}\\
&\rightarrow&\frac{1}{\sqrt{2^{n}}}\sum\limits_{m}(-1)^{x^{'}\cdot m}|m\rangle_{\Rmnum{1}}|s_{A}(m)\oplus r_{A}\rangle_{\Rmnum{2}},
\end{eqnarray}
and sends the two registers to Bob. For the adversary, the mixed state in the channel is:
\begin{eqnarray}\nonumber
\rho&=&\frac{1}{N}\sum\limits_{s_{A}, r_{A}}\sum\limits_{m,n} \frac{(-1)^{m\cdot x^{'}\oplus n\cdot x^{'}}}{2^{l}}|m\rangle_{\Rmnum{1}}\langle n|\otimes |s_{A}(m)\oplus r_{A}\rangle_{\Rmnum{2}}\langle s_{A}(n)\oplus r_{A}|\\
&=&\frac{1}{N2^{l}}\sum\limits_{m, n} (-1)^{m\cdot x^{'}\oplus n\cdot x^{'}}|m\rangle_{\Rmnum{1}}\langle n|\otimes \sum\limits_{s_{A}, r_{A}}|s_{A}(m)\oplus r_{A}\rangle_{\Rmnum{2}}\langle s_{A}(n)\oplus r_{A}|~~
\end{eqnarray}
Up till now, we cannot prove its security yet, though there is no effective attack found.

\textbf{B. Two-round protocol}

Two-round protocol is that includes only the first step of the 4-round protocol. Since $D(\rho_{i}(x), \rho_{j}(y))=0$, information-theoretic security can be guaranteed. Because $D(\rho_{i}(x, s_{A}(\cdot)), \rho_{j}(y, s^{'}_{A}(\cdot))=0$, $D(\rho_{i}(x, s_{B}(\cdot)), \rho_{j}(y, s^{'}_{B}(\cdot))=0$ authentication key $s_{A}(\cdot)$ and $s_{B}(\cdot)$ can be used permanently. However, in this protocols, Alice cannot identify whether the message has been send to Bob. As soon as Bob adds authentication message in the quantum state, the protocol is no longer of provable information-theoretic security, and lost the permanent reusable property of authentication key. The reason is that because that the protocol cannot keep conditions: ${\scriptsize D(\rho_{i}(x), \rho_{j}(y))=0}$, ${\footnotesize D(\rho_{i}(x, s_{A}(\cdot)), \rho_{j}(y, s^{'}_{A}(\cdot))=0}$ in that case.

\textbf{C. Three-round protocol}

Three-round protocol can overcome the difficulty of the two-round protocol described above, and realize mutual authentication, but it cannot ensure the permanent employment of authentication key. To satisfy the security requirement mentioned above, the protocols introduced here compute random Boolean permutation $\{F_{A}(\cdot), F_{B}(\cdot)\}$ controlled by local random numbers to produce entangled states, and use local random numbers $r_{A}$, $r_{B}$ to protect authentication keys $s_{A}(\cdot)$, $s_{B}(\cdot)$. It can be seen that the three-round protocol cannot satisfy all these requirements.

In a three round protocol, the entangled state without message has to be sent by Bob firstly. Then, Alice adds the message in the entangled state in the way as in two-round protocol. As the security analysis in two-round protocol, Alice cannot verify the legitimacy of Bob without consuming key, any further authentication depends on the additional third round. It is difficult for the third round to do that relies on entangled state which is produced by local random number, so the leakage of authentication key is inevitable. Therefore, three-round protocol cannot ensure the permanent use of authentication key while guarantees perfect encryption and origin data authentication.
\section{Conclusion}
We propose a new kind of quantum no-key protocol with provable information-theoretic security. We simplify the initial 9-round protocol to a 4-round protocol, and shows that any protocol of this kind cannot ensure both the security and the permanent employment of the authentication key if its number of rounds is less than or equal to 3.
\section*{Acknowledgement}
This work was supported by the National Natural Science Foundation of China under Grant No.61173157.


\begin{thebibliography}{}

 \bibitem{YW02} Yang L, Wu L -A. Transmit classical and quantum information secretly.
  arXiv: quant-ph/0203089, 2002.
   \bibitem{YWL02}Yang L, Wu L A and Liu S H. A quantum three-pass cryptography protocol. \emph{in Quabtun Optics in Computing and Communication}, SPIE, 2002.
  \bibitem{BF02} Bostr\"{o}m K and Felbinger T. Deterministic secure direct communication using entanglement.
  \emph{Physical Review Letters}, \textbf{89}(18): 187902, 2002.

     \bibitem{Y03} Yang L. Quantum no-key protocol for direct and secure transmission of quantum and classical messages. arXiv: quant-ph/0309200, 2003.
\bibitem{YL12} Yang L, Liang M, Li B, Hu L and Wu L -A. Quantum no-key protocols for secret transmission of quantum and classical message. arXiv: 1112.0981, 2011.
\bibitem{YL121} Yang L and Liang M. Cryptography based on operator theory (\Rmnum{1}): quantum no-key protocols. arXiv:1210.8251, 2012.
\bibitem{JL12}Lang J. A no-key-exchange secure image sharing scheme based on
Shamir¡¯s three-pass cryptography protocol and the multiple-parameter
fractional Fourier transform. \emph{Optics Express}. \textbf{20}(3): 2386-2398, 2012.
 \bibitem{KSM05} Kanamori Y, Yoo S -M and Al-Shurman M. A quantum no-key protocol for secure data communication, in 43rd ACM SE Conference, 2005.
 \bibitem{K06} Kak S. A three stage quantum cryptography protocol. \emph{Foundations of Physics Letters}, \textbf{19}(3): 293, 2006.
 \bibitem{K05} Kye W -H et al. Quantum key distribution with blind polarization bases. \emph{Physical Review Letters}, \textbf{95}(4): 040501, 2005.
 \bibitem{FLS03} Deng F -G , Long G L and Liu X -S. Two-step quantum direct communication protocol using the Einstein-podolsky-Rosen pair block. \emph{Phys. Rev. \textbf{A}}, \textbf{68}(4): 042317, 2003.


\end{thebibliography}
\end{document}